\documentstyle[aps,pra,twocolumn,epsfig]{revtex}

\def\dsum{\mathop{\displaystyle \sum }}

\begin{document}
\title{Physical implementation for entanglement purification of Gaussian continuous
variable quantum states}
\author{Lu-Ming Duan$^{1,2}$\thanks{%
Email: luming.duan@uibk.ac.at}, G. Giedke$^1$, J. I. Cirac$^1$, and P. Zoller%
$^1$}
\address{$^{1}$Institut f\"{u}r Theoretische Physik, Universit\"{a}t Innsbruck,
A-6020 Innsbruck, Austria \\
$^{2}$Department of Physics, University of Science and 
Technology of China, Hefei 230026, China}
\maketitle

\begin{abstract}
We give a detailed description of the entanglement purification protocol
which generates maximally entangled states with high efficiencies from
realistic Gaussian continuous variable entangled states. The physical
implementation of this protocol is extensively analyzed using high finesse
cavities and cavity enhanced cross Kerr nonlinearities. In particular, we
take into account many imperfections in the experimental scheme and
calculate their influences. Quantitative requirements are given for the
relevant experimental parameters.

\end{abstract}

\section{Introduction}

Quantum entanglement plays an essential role in many interesting quantum
information protocols, such as in quantum key distribution and quantum
teleportation \cite{1}. To faithfully realize these protocols, first we need
to generate a maximally entangled state. In reality, however, due to loss
and decoherence, normally we can only generate partially entangled states
between distant sides \cite{2}. Entanglement purification is further needed
which distills a maximally entangled state from several pairs of partially
entangled states using local quantum operations and classical communications 
\cite{3,4}. For qubit systems, efficient entanglement purification protocols
have been found \cite{4,5}. Recently, quantum information protocols have
been extended from qubit systems to continuous variable systems, such as
continuous variable teleportation \cite{6,7}, continuous variable
computation \cite{8} and error correction \cite{11}, continuous variable
cryptography \cite{9}, and also the notions of continuous variable
inseparability \cite{DS} and bound entanglement \cite{HL} have been
investigated. For physical implementation, Gaussian continuous variable
entangled states (i.e., states whose Wigner functions are Gaussians) can be
generated experimentally by transmitting two-mode squeezed light, and this
kind of entanglement has been demonstrated in the recent experiment of
continuous variable teleportation \cite{10}. Obviously, it is useful to
consider purification of continuous variable entanglement, that is, to
generate a desired more entangled state from some realistic continuous
entangled states. We have recently proposed an efficient continuous variable
entanglement purification protocol \cite{14}. In this paper, we present the
mathematical details of this purification protocol together with new results
on its physical implementation. In particular, we take into account many
important imperfections in a realistic experimental setup, and calculate
their influence on the purification scheme. Quantitative requirements are
given for the relevant experimental parameters. These calculations make
necessary preparations for a real experiment. We also show how to generate
Gaussian continuous entangled states between two distant high finesse
cavities, which is the first step for the physical implementation of the
purification protocol.

It should be noted that with direct extensions of the purification protocols
for qubit systems, it is possible to increase entanglement for a special
class of less realistic continuous entangled states \cite{12}.
Unfortunately, with these direct extensions no entanglement increase has
been found till now for realistic Gaussian continuous entangled states. In 
\cite{13} a protocol to increase the entanglement for the special case of
pure two-mode squeezed states has been proposed, which is based on
conditional photon subtraction. For its practical realization, the
efficiency, however, seems to be an issue. In contrast, the purification
scheme discussed in this paper has the following favorable properties: (i)
For pure states it reaches the maximal allowed efficiency in the asymptotic
limit (when the number of pairs of modes goes to infinity); (ii) It can be
readily extended to distill maximally entangled states from a relevant class
of mixed Gaussian states which result from losses in the light transmission;
(iii) An experimental scheme is possible for physical implementation of the
purification protocol using high finesse cavities and cross Kerr
nonlinearities.

The paper is arranged as follows: In section 2 we show how to generate a
Gaussian continuous entangled state between two distant cavities from the
broadband squeezed light provided by a nondegenerate optical parametric
amplifier (NOPA). Light transmission loss is taken into account. In sections
3 and 4 we give a detailed description of the purification protocol. Section
3 shows how to generate a maximally entangled state from pure two-mode
squeezed states based on a local quantum non-demolition (QND) measurement of
the total photon number, and section 4 extends the purification protocol to
include the mixed Gaussian continuous states which are evolved from the pure
two-mode squeezed states due to the unavoidable light transmission loss. In
section 5, we describe a cavity scheme to realize the local QND measurement
of the total photon number, and deduce conditions for the QND\ measurement.
Then, in section 6, we extensively discuss many imperfections for a real
experiment on QND\ measurements, and deduce quantitative requirements for
the relevant experimental parameters. Last, we summarize the results, and
give some typical parameter estimations.

\section{Generation of continuous entangled states between two distant
cavities}

Our source of entangled light field is taken to be a NOPA operating below
threshold \cite{15}. The light fields may be nondegenerate in polarization
or in frequency. The two NOPA cavity modes $c_{A}$ and $c_{B}$ are assumed
to have the same output coupling rate $\kappa _{c}$. The dynamic in the NOPA
cavity is described by the Langevin equations (in the rotating frame) \cite
{16} 
\begin{eqnarray}
\stackrel{.}{c}_{A} &=&\epsilon c_{B}^{\dagger }-\frac{\kappa _{c}}{2}c_{A}-%
\sqrt{\kappa _{c}}c_{iA},  \nonumber \\
\stackrel{.}{c}_{B}^{\dagger } &=&\epsilon ^{\ast }c_{A}-\frac{\kappa _{c}}{2%
}c_{B}^{\dagger }-\sqrt{\kappa _{c}}c_{iB}^{\dagger },  \eqnum{1}
\end{eqnarray}
where $\epsilon $ is the pumping rate with $\left| \epsilon \right| <\kappa
_{c}/2$ (below threshold), and $c_{iA}$ and $c_{iB}$ are vacuum inputs. The
NOPA outputs $c_{oA}$ and $c_{oB}$ are given respectively by $c_{o\alpha
}=c_{i\alpha }+\sqrt{\kappa _{c}}c_{\alpha }$ $\left( \alpha =A,B\right) $.
The two outputs, perhaps after a long distance propagation, are incident on
distant high finesse cavities A and B. The cavities A and B are assumed to
have the same damping rate $\kappa $ with $\kappa \ll \kappa _{c}$. The
schematic setup is shown by Fig. 1.

\begin{figure}[tbp]
\epsfig{file=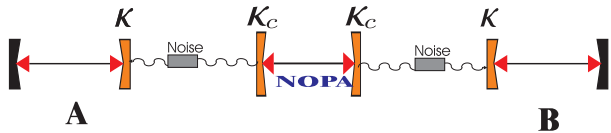,width=8cm}
\caption{Schematic setup for generating Gaussian continuous entangled states
between two distant cavities. }
\end{figure}

Under the condition $\kappa \ll \kappa _{c}$, the dynamics in the NOPA
cavity is much faster than those in the cavities A and B, so we can assume a
steady state for the NOPA outputs. The steady NOPA outputs are described by
squeezed white noise operators with the following correlations \cite{16} 
\begin{eqnarray}
\left\langle c_{oA}\left( t\right) c_{oB}\left( t^{\prime }\right)
\right\rangle  &=&M\delta \left( t-t^{\prime }\right) ,  \nonumber \\
\left\langle c_{o\alpha }^{\dagger }\left( t\right) c_{o\alpha }\left(
t^{\prime }\right) \right\rangle  &=&N\delta \left( t-t^{\prime }\right) ,%
\text{ }\left( \alpha =A,B\right) ,  \eqnum{2}  \label{1} \\
\left\langle c_{o\alpha }\left( t\right) c_{o\alpha }^{\dagger }\left(
t^{\prime }\right) \right\rangle  &=&\left( N+1\right) \delta \left(
t-t^{\prime }\right) ,\text{ }\left( \alpha =A,B\right) ,  \nonumber
\end{eqnarray}
where $N$ and $M$, satisfying $M=\sqrt{N\left( N+1\right) }$, are determined
by the NOPA coupling and pumping rates through $N=\left| \epsilon \right|
^{2}\kappa _{c}^{2}/\left( \frac{\kappa _{c}^{2}}{4}-\left| \epsilon \right|
^{2}\right) ^{2}$ and $M=\left| \epsilon \right| \kappa _{c}\left( \frac{%
\kappa _{c}^{2}}{4}+\left| \epsilon \right| ^{2}\right) /\left( \frac{\kappa
_{c}^{2}}{4}-\left| \epsilon \right| ^{2}\right) ^{2}$.

To get the steady state of the cavities A and B, we note that their inputs $%
a_{iA}$ and $a_{iB}$ are respectively the NOPA outputs $c_{oA}$ and $c_{oB}$
with neglect of the losses during light propagation. The Langevin equations
for the cavity modes $a_{A}$ and $a_{B}$ have the form 
\[
\stackrel{.}{a}_{\alpha }=-\frac{\kappa }{2}a_{\alpha }-\sqrt{\kappa }%
a_{i\alpha },\text{ }\left( \alpha =A,B\right) ,
\]
with the following solution 
\begin{equation}
a_{\alpha }\left( t\right) =a_{\alpha }\left( 0\right) e^{-\frac{\kappa }{2}%
t}-\sqrt{\kappa }\int_{0}^{t}e^{-\frac{\kappa }{2}\left( t-t^{\prime
}\right) }a_{i\alpha }\left( t^{\prime }\right) dt^{\prime }.  \eqnum{3}
\label{3}
\end{equation}
When $\kappa t$ is considerably larger than $1$, from Eqs. (2) and (3), it
follows that 
\begin{eqnarray}
\left\langle a_{A}a_{B}\right\rangle  &=&\sqrt{N\left( N+1\right) }, 
\nonumber \\
\left\langle a_{\alpha }^{\dagger }a_{\alpha }\right\rangle  &=&N,\text{ }%
\left( \alpha =A,B\right) ,  \eqnum{4}  \label{4} \\
\left\langle a_{\alpha }a_{\alpha }^{\dagger }\right\rangle  &=&\left(
N+1\right) ,\text{ }\left( \alpha =A,B\right) .  \nonumber
\end{eqnarray}
On the other hand, we know that two modes driven by a white noise are in
Gaussian states at any time. A Gaussian state with the correlations (4) is
necessarily a pure two-mode squeezed state. So the steady state of the
cavity modes $a_{A}$ and $a_{B}$ is 
\begin{equation}
\left| \Psi \right\rangle _{12}=S_{AB}\left( r\right) \left| \text{vac}%
\right\rangle _{AB},  \eqnum{5}  \label{5}
\end{equation}
where the squeezing operator $S_{AB}\left( r\right) =\exp \left[ r\left(
a_{A}^{\dagger }a_{B}^{\dagger }-a_{A}a_{B}\right) \right] $ and the
squeezing parameter $r$ is determined by $\coth (r)=\sqrt{N+1}.$

Next we include some important sources of noise in the state generation
process. The noise includes the losses in the NOPA cavity and the light
transmission loss from the NOPA\ cavity to the cavities A and B. With a
small loss rate $\eta _{0}\ll \kappa _{c}$ for the modes $c_{A}$ and $c_{B}$
in the NOPA\ cavity, the Langevin equation (1) is replaced by 
\begin{eqnarray}
\stackrel{.}{c}_{A} &=&\epsilon c_{B}^{\dagger }-\frac{\kappa _{c}+\eta _{0}%
}{2}c_{A}-\sqrt{\kappa _{c}}c_{iA}-\sqrt{\eta _{0}}v_{iA},  \nonumber \\
\stackrel{.}{c}_{B}^{\dagger } &=&\epsilon ^{\ast }c_{A}-\frac{\kappa
_{c}+\eta _{0}}{2}c_{B}^{\dagger }-\sqrt{\kappa _{c}}c_{iB}^{\dagger }-\sqrt{%
\eta _{0}}v_{iB}^{\dagger },  \eqnum{6}
\end{eqnarray}
where $v_{iA}$ and $v_{iB}$ are standard vacuum white noise, and the NOPA\
outputs are still given by $c_{o\alpha }=c_{i\alpha }+\sqrt{\kappa _{c}}%
c_{\alpha }$ $\left( \alpha =A,B\right) $. On the other hand, the
transmission loss of light can be described by
\begin{equation}
a_{i\alpha }=c_{o\alpha }\sqrt{e^{-\eta _{\alpha }\tau }}+v_{\alpha }\sqrt{%
1-e^{-\eta _{\alpha }\tau }},\text{ }\left( \alpha =A,B\right) ,  \eqnum{7}
\label{6}
\end{equation}
where $\tau $ is the transmission time, $\eta _{A}$ and $\eta _{B}$ are
respectively the transmission loss rates for the outputs $c_{oA}$ and $c_{oB}
$, and $v_{A}$ and $v_{B}$ are standard vacuum white noise. From Eqs. (6)
and (7), it follows that the inputs for the cavities A and B have the
following correlations

\begin{eqnarray*}
\left\langle a_{iA}\left( t\right) a_{iB}\left( t^{\prime }\right)
\right\rangle  &=&\sqrt{N^{\prime }\left( N^{\prime }+1\right) }e^{-\frac{%
\eta _{A}^{\prime }+\eta _{B}^{\prime }}{2}\tau }\delta \left( t-t^{\prime
}\right) , \\
\left\langle a_{i\alpha }^{\dagger }\left( t\right) a_{i\alpha }\left(
t^{\prime }\right) \right\rangle  &=&N^{\prime }e^{-\eta _{\alpha }^{\prime
}\tau }\delta \left( t-t^{\prime }\right) ,\text{ }\left( \alpha =A,B\right)
, \\
\left\langle a_{i\alpha }\left( t\right) a_{i\alpha }^{\dagger }\left(
t^{\prime }\right) \right\rangle  &=&\left( N^{\prime }e^{-\eta _{\alpha
}^{\prime }\tau }+1\right) \delta \left( t-t^{\prime }\right) ,\text{ }%
\left( \alpha =A,B\right) .
\end{eqnarray*}
where the total loss rates $\eta _{\alpha }^{\prime }=\eta _{\alpha }+\frac{1%
}{\tau }\ln \left( 1+\eta _{0}/\kappa _{c}\right) =\eta _{\alpha }+\eta
_{0}/\left( \kappa _{c}\tau \right) $ $\left( \alpha =A,B\right) $, and the
parameter $N^{\prime }=\left| \epsilon \right| ^{2}\left( \kappa _{c}+\eta
_{0}\right) ^{2}/\left( \frac{\left( \kappa _{c}+\eta _{0}\right) ^{2}}{4}%
-\left| \epsilon \right| ^{2}\right) ^{2}\approx N$. The steady state of the
two cavity modes $a_{A}$ and $a_{B}$ is thus a Gaussian state with the
non-zero correlations given by 
\begin{eqnarray}
\left\langle a_{A}a_{B}\right\rangle  &=&\sqrt{N\left( N+1\right) }e^{-\frac{%
\eta _{A}^{\prime }+\eta _{B}^{\prime }}{2}\tau },  \nonumber \\
\left\langle a_{\alpha }^{\dagger }a_{\alpha }\right\rangle  &=&Ne^{-\eta
_{\alpha }^{\prime }\tau },\text{ }\left( \alpha =A,B\right) ,  \eqnum{8}
\label{8} \\
\left\langle a_{\alpha }a_{\alpha }^{\dagger }\right\rangle  &=&\left(
Ne^{-\eta _{\alpha }^{\prime }\tau }+1\right) ,\text{ }\left( \alpha
=A,B\right) .  \nonumber
\end{eqnarray}
The Gaussian state is completely determined by these correlations. The
Gaussian state (8) can be equivalently described as the solution at time $%
t=\tau $ of the following master equation 
\begin{eqnarray}
\stackrel{.}{\rho } &=&\eta _{A}^{\prime }\left( a_{A}\rho a_{A}^{\dagger }-%
\frac{1}{2}a_{A}^{\dagger }a_{A1}\rho -\frac{1}{2}\rho a_{A}^{\dagger
}a_{A}\right)   \nonumber \\
&&+\eta _{B}^{\prime }\left( a_{B}\rho a_{B}^{\dagger }-\frac{1}{2}%
a_{B}^{\dagger }a_{B}\rho -\frac{1}{2}\rho a_{B}^{\dagger }a_{B}\right)  
\eqnum{9}
\end{eqnarray}
with the initial state $\rho \left( 0\right) =\left| \Psi \right\rangle
_{AB}\left\langle \Psi \right| $, where $\left| \Psi \right\rangle _{AB}$ is
defined by Eq. (5). This equivalence simplifies the physical picture in
section 4, where we will use the master equation (9) to describe the state
generation noise.

\section{Entanglement concentration of pure two-mode squeezed states}

In the above, we have shown how to generate continuous partially entangled
states between two distant cavities. In the case of no noise in the state
generation process, the cavities are in a pure two-mode squeezed state. In
this section, we will show how to concentrate continuous variable
entanglement, that is, starting from several pairs of continuous entangled
states, we want to get a state with more entanglement through only local
operations. The section is divided into two parts. The first part describes
the purification protocol for two entangled pairs, and the second part
extends the protocol to include multiple pairs.

\subsection{Concentration of two entangled pairs}

Assume now we have two cavities $A_{1},A_{2}$ and $B_{1},B_{2}$ on each
side. Each pair of cavities $A_{i},B_{i}$ $\left( i=1,2\right) $ are
prepared in the state (5), which is now denoted by $\left| \Psi
\right\rangle _{A_{i}B_{i}}$. $\left| \Psi \right\rangle _{A_{i}B_{i}}$,
expressed in the number basis, has the form 
\begin{equation}
\left| \Psi \right\rangle _{A_{i}B_{i}}=\sqrt{1-\lambda ^{2}}\stackrel{%
\infty }{\mathrel{\mathop{\sum }\limits_{n=0}}}\lambda ^{n}\left|
n\right\rangle _{A_{i}}\left| n\right\rangle _{B_{i}},  \eqnum{10}
\label{10}
\end{equation}
where $\lambda =\tanh \left( r\right) $. Equation (10) is just the Schmidt
decomposition of the state $\left| \Psi \right\rangle _{A_{i}B_{i}}$. For a
pure state, the entanglement is uniquely quantified by the von Neumann
entropy of the reduced density operator of its one-component. The
entanglement of the state (10) is thus expressed as 
\begin{equation}
E\left( \left| \Psi \right\rangle _{A_{i}B_{i}}\right) =\cosh ^{2}r\log
\left( \cosh ^{2}r\right) -\sinh ^{2}r\log \left( \sinh ^{2}r\right) . 
\eqnum{11}  \label{11}
\end{equation}
The joint state of the two entangled pairs $A_{1},B_{1}$ and $A_{2},B_{2}$
is simply the product 
\begin{eqnarray}
\left| \Psi \right\rangle _{A_{1}B_{1}A_{2}B_{2}} &=&S_{A_{1}B_{1}}\left(
r\right) \left| \text{vac}\right\rangle _{A_{1}B_{1}}\otimes
S_{A_{2}B_{2}}\left( r\right) \left| \text{vac}\right\rangle _{A_{2}B_{2}} 
\nonumber \\
&=&\left( 1-\lambda ^{2}\right) \stackrel{\infty }{\mathrel{\mathop{\sum
}\limits_{j=0}}}\lambda ^{j}\sqrt{1+j}\left| j\right\rangle
_{A_{1}A_{2}B_{1}B_{2}},  \eqnum{12}
\end{eqnarray}
where $\left| j\right\rangle _{A_{1}A_{2}B_{1}B_{2}}$ is defined as 
\begin{equation}
\left| j\right\rangle _{A_{1}A_{2}B_{1}B_{2}}=\frac{1}{\sqrt{1+j}}\stackrel{j%
}{\mathrel{\mathop{\sum }\limits_{n=0}}}\left| n,j-n\right\rangle
_{A_{1}A_{2}}\left| n,j-n\right\rangle _{B_{1}B_{2}}.  \eqnum{13}  \label{13}
\end{equation}

We now perform a local QND measurement of the total photon number of the two
cavities $A_{1},A_{2}$. There have been several proposals for doing QND
measurements of the photon number, and in section 5, we will describe a
cavity scheme for realizing the QND measurement of the total photon number
of two local cavities. Here we simply assume this type of measurement can be
done. After the QND measurement of the total number $n_{A_{1}}+n_{A_{2}}$,
the state $\left| \Psi \right\rangle _{A_{1}B_{1}A_{2}B_{2}}$ is collapsed
into $\left| j\right\rangle _{A_{1}A_{2}B_{1}B_{2}}$ with probability 
\begin{equation}
p_{j}=\left( 1-\lambda ^{2}\right) ^{2}\lambda ^{2j}\left( j+1\right) . 
\eqnum{14}  \label{14}
\end{equation}
The state $\left| j\right\rangle _{A_{1}A_{2}B_{1}B_{2}}$ is a maximally
entangled state between the two parties $A_{1},A_{2}$ and $B_{1},B_{2}$ in a 
$\left( j+1\right) \times \left( j+1\right) $-dimensional Hilbert space, and
its entanglement is 
\begin{equation}
E\left( \left| j\right\rangle _{A_{1}A_{2}B_{1}B_{2}}\right) =\log \left(
j+1\right) .  \eqnum{15}  \label{15}
\end{equation}
If $E\left( \left| j\right\rangle _{A_{1}A_{2}B_{1}B_{2}}\right) >E\left(
\left| \Psi \right\rangle _{A_{i}B_{i}}\right) $, i.e. $j>\frac{\left( \cosh
\left( r\right) \right) ^{\cosh \left( r\right) }}{\left( \sinh \left(
r\right) \right) ^{\sinh \left( r\right) }}-1$, we get a two-party state
with more entanglement. The quantity $\Gamma _{j}=\frac{E\left( \left|
j\right\rangle _{A_{1}A_{2}B_{1}B_{2}}\right) }{E\left( \left| \Psi
\right\rangle _{A_{i}B_{i}}\right) }$ defines the entanglement increase
ratio. Fig. 2 shows the probability of success versus entanglement increase
ratio for some typical values of the squeezing parameter. 
\begin{figure}[tbp]
\epsfig{file=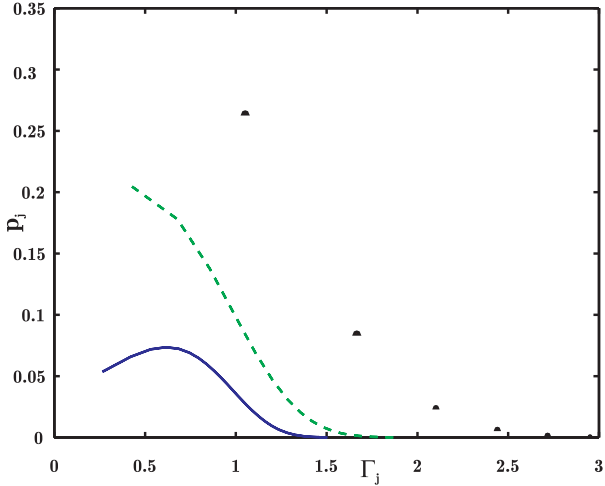,width=8cm}
\caption{The purification success probability versus entanglement increase
ratio for two pairs. Dotted line for the squeezing parameter $r=0.5$, dashed
line for $r=1.0$, and solid line for $r=1.5$. }
\end{figure}

An interesting feature of this entanglement purification protocol is that
with any measurement outcome $j\neq 0$, we always get a useful maximally
entangled state in some finite Hilbert space, though the entanglement of the
outcome state $\left| j\right\rangle _{A_{1}A_{2}B_{1}B_{2}}$ does not
necessarily exceed that of the original state $\left| \Psi \right\rangle
_{A_{i}B_{i}}$ if $j$ is small. The state $\left| j\right\rangle
_{A_{1}A_{2}B_{1}B_{2}}$ involves two pairs of cavities. If one wants to
transfer the entanglement to a single pair of cavity modes, one can make a
phase measurement of the cavity mode $A_{2}$. There have been some proposals
for doing a phase measurement \cite{17,18}. A phase measurement of the mode $%
A_{2}$ with the measurement outcome $\phi $ will convert the state $\left|
j\right\rangle _{A_{1}A_{2}B_{1}B_{2}}$ to the following maximally entangled
state of a single pair of cavity modes 
\begin{equation}
\left| j\right\rangle _{A_{1}A_{2}}=\frac{1}{\sqrt{1+j}}\stackrel{j}{%
\mathrel{\mathop{\sum }\limits_{n=0}}}e^{i\left( j-n\right) \phi }\left|
n\right\rangle _{A_{1}}\left| n\right\rangle _{B_{1}}.  \eqnum{16}
\label{16}
\end{equation}

\subsection{Concentration of multiple entangled pairs}

The above protocol can be extended straightforwardly to simultaneously
concentrate entanglement of multiple cavity-pairs. Simultaneous
concentration of multiple entangled pairs is much more effective that the
entanglement concentration two by two. Assume that we have $m$ cavity-pairs $%
A_{1},B_{1}$, $A_{2},B_{2}$, $\cdots $ and $A_{m},B_{m}$. Each pair of
cavities $A_{i},B_{i}$ is prepared in the state (10). The joint state of the 
$m$ entangled pairs can be expressed as 
\begin{eqnarray}
\left| \Psi \right\rangle _{\left( A_{i}B_{i}\right\} } &=&\left| \Psi
\right\rangle _{A_{1}B_{1}}\otimes \left| \Psi \right\rangle
_{A_{2}B_{2}}\otimes \cdots \otimes \left| \Psi \right\rangle _{A_{m}B_{m}} 
\nonumber \\
&=&\left( 1-\lambda ^{2}\right) ^{\frac{m}{2}}\stackrel{\infty }{%
\mathrel{\mathop{\sum }\limits_{j=0}}}\lambda ^{j}\sqrt{f_{j}^{\left(
m\right) }}\left| j\right\rangle _{\left( A_{i}B_{i}\right\} },  \eqnum{17}
\end{eqnarray}
where $\left( A_{i}B_{i}\right\} $ is abbreviation of$\ A_{1},B_{1}$, $%
A_{2},B_{2},$ $\cdots ,$ $A_{m},B_{m}$, and the normalized state $\left|
j\right\rangle _{\left( A_{i}B_{i}\right\} }$ is defined as 
\begin{equation}
\left| j\right\rangle _{\left( A_{i}B_{i}\right\} }=\frac{1}{\sqrt{%
f_{j}^{\left( m\right) }}}\mathrel{\mathop{\dsum }\limits_{ \begin{array}{c}
i_1,i_2,\cdots ,i_m \\ i_1+i_2+\cdots +i_m=j \end{array} }}\left|
i_{1},i_{2},\cdots ,i_{m}\right\rangle _{\left( A_{i}\right\} }\otimes
\left| i_{1},i_{2},\cdots ,i_{m}\right\rangle _{\left( B_{i}\right\} } 
\eqnum{18}  \label{19}
\end{equation}
The function $f_{j}^{\left( m\right) }$ in Eqs. (17) and (18) is given by 
\begin{equation}
f_{j}^{\left( m\right) }=\frac{\left( j+m-1\right) !}{j!\left( m-1\right) !}%
=\left( 
\begin{array}{c}
j+m-1 \\ 
m-1
\end{array}
\right) .  \eqnum{19}  \label{20}
\end{equation}
To concentrate the entanglement, we perform a QND measurement of the total
photon number $n_{A_{1}}+n_{A_{2}}+\cdots +n_{A_{m}}$. This measurement
projects the state $\left| \Psi \right\rangle _{\left( A_{i}B_{i}\right\} }$
onto a two-party maximally entangled state $\left| j\right\rangle _{\left(
A_{i}B_{i}\right\} }$ with probability 
\begin{equation}
p_{j}^{\left( m\right) }=\left( 1-\lambda ^{2}\right) ^{m}\lambda
^{2j}f_{j}^{\left( m\right) }.  \eqnum{20}  \label{21}
\end{equation}
The entanglement of the outcome state $\left| j\right\rangle _{\left(
A_{i}B_{i}\right\} }$ is given by 
\begin{equation}
E\left( \left| j\right\rangle _{\left( A_{i}B_{i}\right\} }\right) =\log
\left( f_{j}^{\left( m\right) }\right) .  \eqnum{21}  \label{22}
\end{equation}
Similarly, $\Gamma _{j}=E\left( \left| j\right\rangle _{\left(
A_{i}B_{i}\right\} }\right) /E\left( \left| \Psi \right\rangle
_{A_{i}B_{i}}\right) $ defines the entanglement increase ratio, and if $%
\Gamma _{j}>1$, we get a more entangled state. For four pairs, the
probability of success versus entanglement increase ratio is shown in Fig.
3. There appears a peak in the probability curve for some entanglement
increase ratio between $2$ and $3$. 
\begin{figure}[tbp]
\epsfig{file=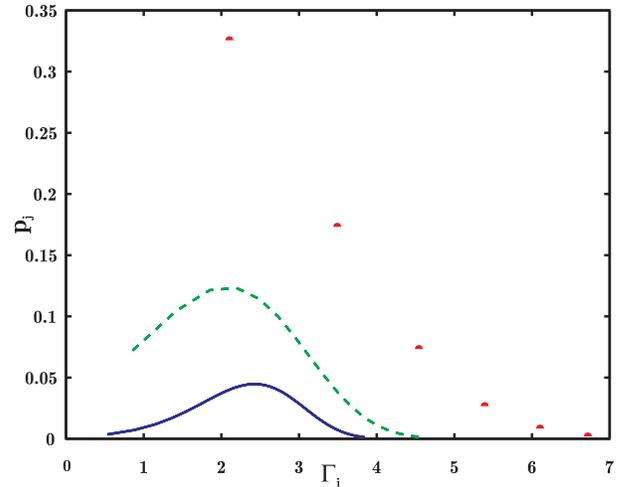,width=8cm}
\caption{The purification success probability versus entanglement increase
ratio for the number of pairs $m=4$. Dotted line for the squeezing parameter 
$r=0.5$, dashed line for $r=1.0$, and solid line for $r=1.5$. }
\end{figure}

To measure how efficient the scheme is, we define the entanglement transfer
efficiency $\Upsilon $ with the expression 
\begin{equation}
\Upsilon =\frac{\stackrel{\infty }{\mathrel{\mathop{\sum }\limits_{j=0}}}%
p_{j}^{\left( m\right) }E\left( \left| j\right\rangle _{\left(
A_{i}B_{i}\right\} }\right) }{mE\left( \left| \Psi \right\rangle
_{A_{i}B_{i}}\right) }.  \eqnum{22}  \label{23}
\end{equation}
It is the ratio of the average entanglement after concentration measurement
to the initial total entanglement contained in the $m$ pairs. Obviously, $%
\Upsilon \leq 1$ should always hold. With the squeezing parameter $r=0.5,$ $%
1.0$ or $1.5,$ the entanglement transfer efficiency versus the number of
pairs $m$ is shown in Fig. 4.

\begin{figure}[tbp]
\epsfig{file=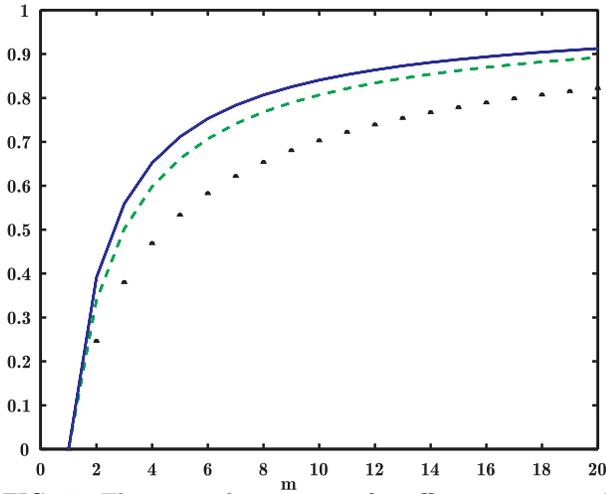,width=8cm}
\caption{The entanglement transfer efficiency versus the number of pairs $m$
in simultaneous concentration. Dotted line for $r=0.5$, dashed line for $%
r=1.0$, and solid line for $r=1.5$. }
\end{figure}
From the figure, we see that the entanglement transfer efficiency is near to
1 for a large number of pairs. In fact, it can be proven that if $m$ goes to
infinity, with unit probability we would get a maximally entangled state
with entanglement $mE\left( \left| \Psi \right\rangle _{A_{i}B_{i}}\right) .$
To show this, we calculate the mean value and the variance of the
distribution $p_{j}^{\left( m\right) }$, and find 
\begin{eqnarray}
\overline{j} &=&\frac{m\lambda ^{2}}{\left( 1-\lambda ^{2}\right) }, 
\nonumber \\
\overline{\left( \Delta j\right) ^{2}} &=&\frac{m\lambda ^{2}}{\left(
1-\lambda ^{2}\right) ^{2}}.  \eqnum{23}
\end{eqnarray}
The results show that if $m$ tends to infinity, $\sqrt{\overline{\left(
\Delta j\right) ^{2}}}/\overline{j}\rightarrow 0$\ and the distribution $%
p_{j}^{\left( m\right) }$ tends to a $\delta $-like function. Furthermore,
around the mean value $\overline{j}$, the entanglement of the resulting
state $\left| \overline{j}\right\rangle _{\left( A_{i}B_{i}\right\} }$ is 
\begin{equation}
E\left( \left| \overline{j}\right\rangle _{\left( A_{i}B_{i}\right\}
}\right) \stackrel{m\rightarrow \infty }{\longrightarrow }mE\left( \left|
\Psi \right\rangle _{A_{i}B_{i}}\right) ,  \eqnum{24}
\end{equation}
so the entanglement transfer efficiency tends to unity. This proves that the
purification method described above is optimal in the asymptotic limit ($%
m\rightarrow \infty $), analogous to the purification protocol presented in 
\cite{4} for the qubit case. For any finite number of entangled pairs, this
purification protocol is more efficient than that in \cite{4}, since it
takes advantage of the special relations between the coefficients in the
two-mode squeezed state.

\section{Entanglement purification of mixed Gaussian continuous entangled
states}

The assumption of noise-free preparation of partially continuous entangled
states is not realistic. If we include the unavoidable light transmission
loss and the NOPA cavity loss in the state generation process, in section 2
we have shown that we would get a mixed Gaussian continuous entangled state
between two distant cavities. The state is described by the solution at the
transmission time $\tau $ of the master equation (9), with the ideal
two-mode squeezed state (10) at the beginning. If we want to establish $m$
entangled cavity-pairs $A_{1},B_{1}$, $A_{2},B_{2}$, $\cdots $ and $%
A_{m},B_{m}$, Eq. (9) can be extended directly to the following form 
\begin{equation}
\stackrel{.}{\rho }=-i\left( H_{\text{eff}}\rho -\rho H_{\text{eff}%
}^{\dagger }\right) +\stackrel{m}{\mathrel{\mathop{\sum }\limits_{i=1}}}%
\left( \eta _{A}^{\prime }a_{A_{i}}\rho a_{A_{i}}^{\dagger }+\eta
_{B}^{\prime }a_{B_{i}}\rho a_{B_{i}}^{\dagger }\right)   \eqnum{25}
\label{24}
\end{equation}
where $\rho $ is the density operator of the whole $m$ entangled pairs with $%
\rho \left( 0\right) =\left| \Psi \right\rangle _{\left( A_{i}B_{i}\right\}
}\left\langle \Psi \right| $, and the effective Hamiltonian 
\begin{equation}
H_{\text{eff}}=-i\stackrel{m}{\mathrel{\mathop{\sum }\limits_{i=1}}}\left( 
\frac{\eta _{A}^{\prime }}{2}a_{A_{i}}^{\dagger }a_{A_{i}}+\frac{\eta
_{B}^{\prime }}{2}a_{B_{i}}^{\dagger }a_{B_{i}}\right) .  \eqnum{26}
\label{25}
\end{equation}
In Eqs. (25) and (26), we assumed that the total loss rates $\eta
_{A}^{\prime }$ and $\eta _{B}^{\prime }$ are the same for the $m$ entangled
pairs, but $\eta _{A}^{\prime }$ and $\eta _{B}^{\prime }$ may be different
from each other. In this section, we will show how to distill entanglement
from the kind of realistic continuous entangled states described by the
solution of the master equation (25). There are two practical circumstances
in which our entanglement purification protocol can be extended
straightforwardly to generate maximally entangled states from the mixed
Gaussian entangled states. We describe these two circumstances one by one.

\subsection{Case of small state preparation noise}

Though the state preparation noise is unavoidable, in many cases it is
reasonable to assume that it is quite small. We take $\eta _{A}^{\prime
}\tau $ and $\eta _{B}^{\prime }\tau $ as small factors, and solve the
master equation (25) perturbatively to the first order of these small
factors. It is convenient to use the quantum trajectory language to explain
the perturbative solution. In this language, to the first order of $\eta
_{A}^{\prime }\tau $ and $\eta _{B}^{\prime }\tau $, the final normalized
state of the $m$ entangled pairs is either (no jumps) 
\begin{eqnarray}
\left| \Psi ^{\left( 0\right) }\right\rangle _{\left( A_{i}B_{i}\right\} }
&=&\frac{1}{\sqrt{p^{\left( 0\right) }}}e^{-iH_{\text{eff}}\tau }\left| \Psi
\right\rangle _{\left( A_{i}B_{i}\right\} }  \nonumber \\
&=&\frac{1}{\sqrt{p^{\left( 0\right) }}}\left( 1-\lambda ^{2}\right) ^{\frac{%
m}{2}}\stackrel{\infty }{\mathrel{\mathop{\sum }\limits_{j=0}}}\lambda
^{j}e^{-\frac{\eta _{A}^{\prime }+\eta _{B}^{\prime }}{2}\tau j}\sqrt{%
f_{j}^{\left( m\right) }}\left| j\right\rangle _{\left( A_{i}B_{i}\right\} },
\eqnum{27}
\end{eqnarray}
with probability 
\begin{equation}
p^{\left( 0\right) }=\frac{\left( 1-\lambda ^{2}\right) ^{m}}{\left(
1-\lambda ^{2}e^{-\left( \eta _{A}^{\prime }+\eta _{B}^{\prime }\right) \tau
}\right) ^{m}}  \eqnum{28}  \label{27}
\end{equation}
or (a jump occurred) 
\begin{equation}
\left| \Psi ^{\left( \alpha _{i}\right) }\right\rangle _{\left(
A_{i}B_{i}\right\} }=\frac{1}{\sqrt{p^{\left( \alpha _{i}\right) }}}\sqrt{%
\eta _{\alpha }^{\prime }\tau }a_{\alpha _{i}}\left| \Psi \right\rangle
_{\left( A_{i}B_{i}\right\} },\text{ }\left( \alpha =A,B\text{ and }%
i=1,2,\cdots ,m\right)   \eqnum{29}  \label{28}
\end{equation}
with probability 
\begin{eqnarray}
p^{\left( \alpha _{i}\right) } &=&\eta _{\alpha }^{\prime }\tau \text{ }%
_{\left( A_{i}B_{i}\right\} }\left\langle \Psi \right| a_{\alpha
_{i}}^{\dagger }a_{\alpha _{i}}\left| \Psi \right\rangle _{\left(
A_{i}B_{i}\right\} }  \nonumber \\
&=&\overline{n}\eta _{\alpha }^{\prime }\tau ,  \eqnum{30}
\end{eqnarray}
where $\overline{n}=_{\left( A_{i}B_{i}\right\} }\left\langle \Psi \right|
a_{\alpha _{i}}^{\dagger }a_{\alpha _{i}}\left| \Psi \right\rangle _{\left(
A_{i}B_{i}\right\} }=\sinh ^{2}\left( r\right) $ is the mean photon number
for a single mode.

Similar to the pure state case, we also use QND\ measurements of the total
photon number to distill entanglement from the mixed continuous state
described by Eqs. (27)-(30). The difference is that now we perform QND
measurements on both sides A and B. The measurement results are denoted by $%
j_{A}$ and $j_{B}$, respectively. We then compare $j_{A}$ and $j_{B}$
through classical communication, and keep the outcome state if and only if $%
j_{A}=j_{B}$. It is easy to show that the final state is a maximally
entangled state in a finite dimensional Hilbert space. Let $P_{A}^{\left(
j\right) }$ and $P_{B}^{\left( j\right) }$ denote the projections onto the
eigenspace of the corresponding total number operator $\stackrel{m}{%
\mathrel{\mathop{\sum
}\limits_{i=1}}}a_{A_{i}}^{\dagger }a_{A_{i}}$ and $\stackrel{m}{%
\mathrel{\mathop{\sum
}\limits_{i=1}}}a_{B_{i}}^{\dagger }a_{B_{i}}$ with eigenvalue $j$,
respectively. From Eqs. (27) and (29), it follows 
\begin{eqnarray}
P_{A}^{\left( j\right) }P_{B}^{\left( j\right) }\left| \Psi ^{\left(
0\right) }\right\rangle _{\left( A_{i}B_{i}\right\} } &=&\left|
j\right\rangle _{\left( A_{i}B_{i}\right\} },  \nonumber \\
P_{A}^{\left( j\right) }P_{B}^{\left( j\right) }\left| \Psi ^{\left( \alpha
_{i}\right) }\right\rangle _{\left( A_{i}B_{i}\right\} } &=&0,\text{ }\left(
\alpha =A,B\text{ and }i=1,2,\cdots ,m\right)   \eqnum{31}
\end{eqnarray}
So if $j_{A}=j_{B}$, the outcome state is maximally entangled with
entanglement $\log \left( f_{j}^{\left( m\right) }\right) $. The components
(29) in the mixed density operator, which are not maximally entangled, are
discarded through confirmation of the two-side measurement outcomes.
Compared with the pure state case, the probability to get the entangled
state $\left| j\right\rangle _{\left( A_{i}B_{i}\right\} }$ is now decreased
to 
\begin{equation}
p_{j}^{\prime }=\left( 1-\lambda ^{2}\right) ^{m}\lambda ^{2j}f_{j}^{\left(
m\right) }e^{-\left( \eta _{A}^{\prime }+\eta _{B}^{\prime }\right) \tau j}.
\eqnum{32}  \label{31}
\end{equation}
We also note that the projection operators $P_{A}^{\left( j\right)
}P_{B}^{\left( j\right) }$ cannot eliminate the state obtained from the
initial state $\left| \Psi \right\rangle _{\left( A_{i}B_{i}\right\} }$ by a
quantum jump on both sides A and B. The total probability for this kind of
quantum jumps to occur is proportional to $m^{2}\overline{n}^{2}\eta
_{A}^{\prime }\eta _{B}^{\prime }\tau ^{2}$. So the condition for small
state preparation noise in fact requires 
\begin{equation}
m^{2}\overline{n}^{2}\left( \eta _{A}\tau +\eta _{0}/\kappa _{c}\right)
\left( \eta _{B}\tau +\eta _{0}/\kappa _{c}\right) \ll 1.  \eqnum{33}
\label{32}
\end{equation}
If the light transmission loss is the dominant noise, Eq. (33) reduces to $%
m^{2}\overline{n}^{2}\eta _{A}\eta _{B}\tau ^{2}\ll 1$.

\subsection{Case of asymmetric state preparation noise}

In the above purification protocol, we need classical communication (CC) to
confirm that the measurement outcomes of the two sides are the same, and
during this CC, we implicitly assume that the storage noise for the cavity
modes is negligible. In fact, that the storage noise during CC is much
smaller than the transmission noise is a common assumption made in all the
entanglement purification schemes which need the help of repeated CCs \cite
{3,5}. If we also make this assumption for continuous variable systems,
there exists a simple purification protocol to generate maximally entangled
states. We put the NOPA setup on the A side. After creation of ideal
squeezed vacuum lights, we directly couple one output light of the NOPA to
the cavity on side A without noisy propagation; and the other output of the
NOPA is sent to the remote side B, through a long distance noisy
transmission. This configuration of the setup is equivalent to setting the
transmission loss rate $\eta _{A}\approx 0$ so that $\eta _{A}^{\prime
}\approx \eta _{0}/\left( \kappa _{c}\tau \right) $. Note that the NOPA
cavity loss rate $\eta _{0}$ is normally much smaller than the output
coupling rate $\kappa _{c}$, so the total loss rate $\eta _{A}^{\prime }$
can be much smaller than $\eta _{B}^{\prime }$ in this case. The
purification protocol now is exactly the same as that described in the
previous case. We note that the component of the mixed density operator
which is kept the projection $P_{A}^{\left( j\right) }P_{B}^{\left( j\right)
}$ should subject to the same times of quantum jumps on each side A and B.
We want this component is a maximally entangled state. This requires that
the total probability for A and B to be subjected to the same nonzero number
of quantum jumps should be very small. From Eq. (30), this total probability
is always smaller than $m\overline{n}\eta _{A}^{\prime }\tau $, no matter
how large the transmission loss $\eta _{B}\tau $ is. So the working
condition of the protocol in the asymmetric transmission noise case is 
\begin{equation}
m\overline{n}\eta _{0}/\kappa _{c}\ll 1.  \eqnum{34}  \label{33}
\end{equation}
The transmission loss $\eta _{B}\tau $ can be above one. The probability of
success for obtaining the maximally entangled state $\left| j\right\rangle
_{\left( A_{i}B_{i}\right\} }$ is also given by Eq. (32).

Before concluding this section, we remark that for continuous variable
systems, the information carrier is normally light, and the assumption of
storage with a very small loss rate is typically unrealistic. It is
interesting to note that recently there have been proposals to store light
in internal states of an atomic ensemble \cite{19,Lukin}. If this turns out
to be possible, the storage time for light can be greatly increased. Anyway,
as was pointed out in \cite{14}, this purification method is in fact not
essentially hampered by the difficulty to store light, since there is a
simple posterior confirmation method to circumvent the storage problem. Note
that the purpose to distill maximally entangled states is to directly apply
them in some quantum communication protocol, such as in quantum cryptography
or in quantum teleportation. So we can modify the above purification
protocol by the following procedure: right after the cavity A attains its
steady state, we make a QND measurement of the total excitation number on
side A and get a measurement result $j_{A}$. Then we do not store the
outcome state on side A, but immediately use it (e.g., perform the
corresponding measurement as required by a quantum cryptography protocol).
During this process, the modes $B_{i}$ are being sent to the distant side B,
and when they arrive, we make another QND measurement of the total
excitation number of the modes $B_{i}$ and get a outcome $j_{B}$. The
resulting state on side B can be directly used (for quantum cryptography for
instance) if $j_{A}=j_{B},$ and discarded otherwise. By this method, we
formally get maximally entangled states through posterior confirmation, and
at the same time we need not store the modes on both sides.

\section{QND measurements of the total photon number of several cavities}

The QND measurement of the total photon number plays a critical role in our
entanglement purification protocol. There have been some proposals for
making a QND\ measurement of the photon number in a single cavity \cite
{20,22,23}, such as letting some atoms pass through the cavity, and
measuring the internal or external degrees of freedom of the atoms \cite{20}%
. In this section, we propose a purely optical scheme for making a QND
measurement of the total photon number contained in several cavities. The
different optical modes interact with each other through cross phase
modulation induced by a Kerr medium, and we use cavities to enhance this
kind of interaction. As an illustrative example, in the following we will
show how to measure the total photon number of two cavities. Extension of
this scheme to include several cavities is straightforward.

The schematic setup is depicted in Fig. 5. We want to make a QND\
measurement of the total photon number $n_{1}+n_{2}$ contained in the good
cavities I and II, whose damping rate $\kappa $ is assumed to be very small.
The cavities I and II, each with a Kerr type medium inside, are put
respectively in a bigger ring cavity. The two ring cavities are assumed to
damp at the same rate $\gamma $, and $\gamma \gg \kappa $. A strong coherent
light $b_{i1}$ is incident on the first ring cavity, whose output $b_{o1}$
is directed to the second ring cavity. The output $b_{o2}$ of the second
ring cavity is continuously observed through homodyne detection, and we will
show that under some realistic conditions, this detection gives a QND
measurement of the total photon number operator $n_{1}+n_{2}=a_{1}^{\dagger
}a_{1}+a_{2}^{\dagger }a_{2}$.

\begin{figure}[tbp]
\epsfig{file=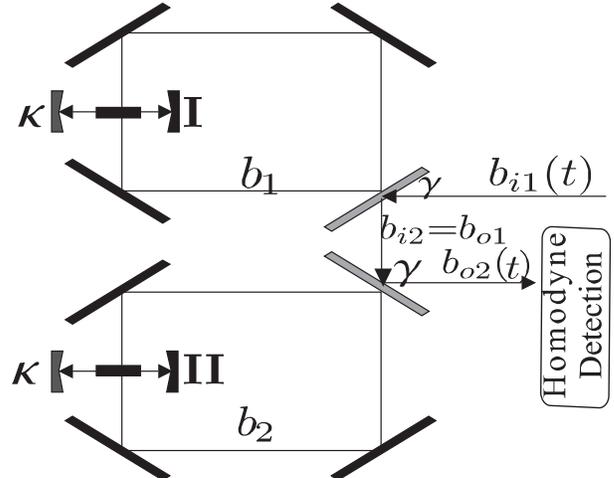,width=8cm}
\caption{Schematic experimental setup to measure the total photon number $%
n_{1}+n_{2}$ contained in the cavities I and II. }
\end{figure}

The measurement model depicted in Fig. 5 is an example of a cascaded quantum
system \cite{16}.The incident light $b_{i1}$ can be expressed as 
\begin{equation}
b_{i1}=b_{i1}^{\prime }+g\sqrt{\gamma },  \eqnum{35}  \label{34}
\end{equation}
where $g\sqrt{\gamma }$ is a constant driving field, and $b_{i1}^{^{\prime
}} $ is the standard vacuum white noise, satisfying 
\begin{eqnarray}
\left\langle b_{i1}^{\prime \dagger }\left( t\right) b_{i1}^{\prime }\left(
t^{\prime }\right) \right\rangle &=&0,  \nonumber \\
\left\langle b_{i1}^{\prime }\left( t\right) b_{i1}^{\prime \dagger }\left(
t^{\prime }\right) \right\rangle &=&\delta \left( t-t^{\prime }\right) . 
\eqnum{36}
\end{eqnarray}
The Hamiltonian for the Kerr medium is assumed to be 
\begin{equation}
H_{i}=\hbar \chi n_{i}b_{i}^{\dagger }b_{i},\text{ }\left( i=1,2\right) , 
\eqnum{37}  \label{36}
\end{equation}
where $b_{1}$ and $b_{2}$ are the annihilation operators for the ring cavity
modes, and $\chi $ is the cross-phase modulation coefficient. The self-phase
modulation effects will be discussed in the next section and shown to be
negligible under some realistic conditions. In the rotating frame, the
Langevin equations describing the dynamics in the two ring cavities have the
form 
\begin{eqnarray}
\stackrel{.}{b}_{1} &=&-i\chi n_{1}b_{1}-\frac{\gamma }{2}b_{1}-\sqrt{\gamma 
}b_{i1}^{\prime }-g\gamma ,  \nonumber \\
\stackrel{.}{b}_{2} &=&-i\chi n_{2}b_{2}-\frac{\gamma }{2}b_{2}-\sqrt{\gamma 
}b_{i2}  \eqnum{38}
\end{eqnarray}
The boundary conditions for the two ring cavities are described by 
\begin{eqnarray}
b_{i2} &=&b_{o1}=b_{i1}^{\prime }+g\sqrt{\gamma }+\sqrt{\gamma }b_{1}, 
\nonumber \\
b_{o2} &=&b_{i2}+\sqrt{\gamma }b_{2}.  \eqnum{39}
\end{eqnarray}
Assume $\gamma \gg \chi \left\langle n_{i}\right\rangle ,$ $\left(
i=1,2\right) $, and we take adiabatic elimination, i.e., let $\stackrel{.}{b}%
_{1}=\stackrel{.}{b}_{2}=0$ in Eq. (38), obtaining 
\begin{eqnarray}
b_{1} &\approx &\frac{-2\left( g\gamma +\sqrt{\gamma }b_{i1}^{\prime
}\right) }{\gamma }\left( 1-\frac{2i\chi n_{1}}{\gamma }\right) ,  \nonumber
\\
b_{2} &\approx &\frac{2\left( g\gamma +\sqrt{\gamma }b_{i1}^{\prime }\right) 
}{\gamma }\left( 1-\frac{4i\chi n_{1}}{\gamma }-\frac{2i\chi n_{2}}{\gamma }%
\right)  \eqnum{40}
\end{eqnarray}
Substituting the above result into Eq. (39), the final output field $b_{o2}$
is expressed as 
\begin{equation}
b_{o2}\approx -\frac{4ig\chi }{\sqrt{\gamma }}\left( n_{1}+n_{2}\right)
+b_{i1}^{\prime }+g\sqrt{\gamma }.  \eqnum{41}  \label{40}
\end{equation}

Now we measure the $X$-component of the quadrature phase amplitudes of the
output field $b_{o2}$ through a homodyne detection. The phase of the driving
field $g$ is set according to $g=i\left| g\right| $. Suppose $T$ is the
measuring time. What we really get is the integrated photon current over
time $T$, which, divided by $T,$ corresponds to the following measuring
operator 
\begin{eqnarray}
X_{T} &=&\frac{1}{T}\int_{0}^{T}\frac{1}{\sqrt{2}}\left[ b_{o2}\left(
t\right) +b_{o2}^{\dagger }\left( t\right) \right] dt  \nonumber \\
&\approx &\frac{4\sqrt{2}\left| g\right| \chi }{\sqrt{\gamma }}\left(
n_{1}+n_{2}\right) +\frac{1}{\sqrt{T}}X_{T}^{\left( b\right) },  \eqnum{42}
\end{eqnarray}
where $X_{T}^{\left( b\right) }=\frac{1}{\sqrt{2}}\left(
b_{T}+b_{T}^{\dagger }\right) $, and $b_{T}$, satisfying $\left[
b_{T},b_{T}^{\dagger }\right] =1$, is defined by 
\begin{equation}
b_{T}=\frac{1}{\sqrt{T}}\int_{0}^{T}b_{i1}^{\prime }\left( t\right) dt. 
\eqnum{43}  \label{42}
\end{equation}
From Eq. (36), it follows that the defined mode $b_{T}$ is in a vacuum
state. So the first term of the right hand side of Eq. (42) represents the
signal which is proportional to $n_{1}+n_{2}$, and the second term
represents the contribution of the vacuum noise. The distinguishability of
this measurement is given by 
\begin{equation}
\delta n=\frac{\sqrt{\gamma }}{8\left| g\right| \chi \sqrt{T}}.  \eqnum{44}
\label{43}
\end{equation}
If $\delta n<1$, i.e., if the measuring time 
\begin{equation}
T>\frac{\gamma }{64\left| g\right| ^{2}\chi ^{2}},  \eqnum{45}  \label{44}
\end{equation}
we perform an effective measurement of the total number operator $%
n_{1}+n_{2} $. During the measuring time $T$, the loss of the two cavities I
and II should be negligible, which requires 
\begin{equation}
\kappa \left\langle n_{i}\right\rangle T<1,\text{ }\left( i=1,2\right) 
\eqnum{46}  \label{45}
\end{equation}
Under this condition, $n_{1}+n_{2}$ is approximately a conserved observable,
and we realize a QND measurement of the total photon number operator$.$ The
measurement projects the field in the cavities I and II to one of the
eigenstates of $n_{1}+n_{2}$. Eqs. (45) and (46), combined together,
determine the suitable choice for the measuring time.

\section{Influence of imperfections in the QND measurement}

We have shown how to perform a QND\ measurement of the total photon number.
The scheme described above works under ideal conditions. For a real
experiment, there are always many imperfections which should be considered.
For example, the phase of the driving field may be unstable, and has a small
variance; the damping rates and the cross phase modulation coefficients for
different ring cavities may not be exactly the same; the Kerr media and the
mirrors may absorb some light; self-phase modulation effects caused by the
Kerr media may have some influence on the resulting state; there may be some
loss of light from the first ring cavity to the second ring cavity; the
efficiency of the detector is not unity. Of course, to realize a QND\
measurement of the total photon number, all the imperfections must be small.
But the important question is how small these imperfections should be. In
this section, we will deduce quantitative requirements for all the
imperfections listed above. These calculations may be helpful for a future
real experiment. We will consider these imperfections one by one.

\subsection{Phase instability of the driving field}

Assume that the phase of the driving field $g\sqrt{\gamma }$ has a small
variance $\delta $, i.e., $g$ is expressed as $g=i\left| g\right| e^{i\delta
}$. Then, Eq. (42) is replaced by 
\begin{equation}
X_{T}\approx \frac{4\sqrt{2}\left| g\right| \chi }{\sqrt{\gamma }}\left(
n_{1}+n_{2}\right) +\frac{1}{\sqrt{T}}X_{T}^{\left( b\right) }-\sqrt{2}%
\left| g\right| \delta \sqrt{\gamma },  \eqnum{47}  \label{46}
\end{equation}
The last term of Eq. (47) represents the noise due to the phase instability
of the driving field. It should be negligible compared with the signal,
which requires 
\begin{equation}
\delta <\frac{4\chi }{\gamma }.  \eqnum{48}  \label{47}
\end{equation}
On the other hand, we know that the squared phase variance $\delta ^{2}$
increases linearly with time, i.e. $\delta ^{2}=\delta _{t}t$, where $\delta
_{t}$ is the increasing rate. The measuring time $T$ is bounded from below
by Eq. (45), so the increasing rate of the phase instability of the driving
field is required to satisfy 
\begin{equation}
\delta _{t}<\frac{1024\left| g\right| ^{2}\chi ^{4}}{\gamma ^{3}}. 
\eqnum{49}  \label{48}
\end{equation}
Eq. (49) suggests it is easier to meet the requirement imposed by the phase
instability with a strong driving field and a large cross phase modulation
coefficient.

\subsection{Imbalance between the ring cavities}

In the previous section, we assumed that the damping rates and the cross
phase modulation coefficients are exactly the same for the two ring
cavities. This may be impossible in a real experiment. Here we calculate the
largest allowed imbalance between the two ring cavities. The damping rates
and the cross phase modulation coefficients for the ring cavities are
denoted by $\gamma _{1},$ $\gamma _{2}$ and $\chi _{1},$ $\chi _{2}$,
respectively. The Langevin equations (38) and the boundary conditions (39)
are replaced respectively by the following equations 
\begin{eqnarray}
\stackrel{.}{b}_{1} &=&-i\chi _{1}n_{1}b_{1}-\frac{\gamma _{1}}{2}b_{1}-%
\sqrt{\gamma _{1}}b_{i1}^{\prime }-g\gamma _{1},  \nonumber \\
\stackrel{.}{b}_{2} &=&-i\chi _{2}n_{2}b_{2}-\frac{\gamma _{2}}{2}b_{2}-%
\sqrt{\gamma _{2}}b_{i2}  \eqnum{50}
\end{eqnarray}
\begin{eqnarray}
b_{i2} &=&b_{o1}=b_{i1}^{\prime }+g\sqrt{\gamma _{1}}+\sqrt{\gamma _{1}}%
b_{1},  \nonumber \\
b_{o2} &=&b_{i2}+\sqrt{\gamma _{2}}b_{2}.  \eqnum{51}
\end{eqnarray}
The final measured observable is expressed as 
\begin{equation}
X_{T}\approx \frac{4\sqrt{2}\left| g\right| \chi _{1}}{\sqrt{\gamma _{1}}}%
\left( n_{1}+n_{2}\right) +\frac{1}{\sqrt{T}}X_{T}^{\left( b\right) }+4\sqrt{%
2}\left| g\right| \sqrt{\gamma _{1}}\left( \frac{\chi _{2}}{\gamma _{2}}-%
\frac{\chi _{1}}{\gamma _{1}}\right) n_{2},  \eqnum{52}  \label{51}
\end{equation}
The last term of Eq. (52) represents the noise due to the unbalance between
the ring cavities, which should be negligible compared with the signal,
yielding 
\begin{equation}
\left| \frac{\chi _{2}\gamma _{1}}{\chi _{1}\gamma _{2}}-1\right| <\frac{1}{%
\left\langle n_{2}\right\rangle }.  \eqnum{53}  \label{52}
\end{equation}

\subsection{Absorption and leakage of the light}

Light absorption by mirrors and Kerr media and light leakage through other
mirrors of the ring cavities can be described by the same Langevin equation,
which has the form 
\begin{eqnarray}
\stackrel{.}{b}_{1} &=&-i\chi n_{1}b_{1}-\frac{\gamma }{2}b_{1}-\sqrt{\gamma 
}b_{i1}^{\prime }-g\gamma -\frac{\beta _{1}}{2}b_{1}-\sqrt{\beta _{1}}c_{i1},
\nonumber \\
\stackrel{.}{b}_{2} &=&-i\chi n_{2}b_{2}-\frac{\gamma }{2}b_{2}-\sqrt{\gamma 
}b_{i2}-\frac{\beta _{2}}{2}b_{2}-\sqrt{\beta _{2}}c_{i2},  \eqnum{54}
\end{eqnarray}
where $\beta _{1}$ and $\beta _{2}$ are the light leakage (or absorption)
rates of the first and second ring cavities, respectively, and $c_{i1}$ and $%
c_{i2}$ are the standard vacuum inputs. The boundary conditions for the ring
cavities are still described by Eq. (39). The leaked (or absorbed) light
fields $c_{o1}$ and $c_{o2}$ are expressed as 
\begin{equation}
c_{o\alpha }=c_{i\alpha }+\sqrt{\beta _{\alpha }}b_{\alpha },\text{ }\left(
\alpha =1,2\right) .  \eqnum{55}  \label{54}
\end{equation}
The leakage (or absorption) of light may have two types of effects: First,
it may destroy the balance between the two ring cavities; and second, the
leaked light (55) may carry some information about $n_{1}$ (or $n_{2}$). Any
information about $n_{1}$ (or $n_{2}$) will destroy the superposition of the
different eigenstates of $n_{1}$ (or $n_{2}$), and thus lead to decoherence
of the eigenstate of $n_{1}+n_{2}$ (Note that a eigenstate of $n_{1}+n_{2}$
is normally a superposition of the different eigenstates of $n_{1}$ (or $%
n_{2}$)). So we require that the information about $n_{1}$ (or $n_{2}$)
carried by the leaked light should be completely masked by the vacuum noise.
This is equivalent to require that the decoherence of the eigenstate of $%
n_{1}+n_{2}$ caused by the light leakage is negligible. To consider the
first effect of the light leakage, we calculate the measured observable $%
X_{T}$, and find it has the form 
\begin{equation}
X_{T}\approx \frac{4\sqrt{2}\left| g\right| \chi }{\sqrt{\gamma }}\left(
n_{1}+n_{2}\right) +\frac{1}{\sqrt{T}}X_{T}^{\left( b\right) }+\frac{4\sqrt{2%
}\left| g\right| \chi }{\sqrt{\gamma }}\left( \frac{\beta _{2}^{2}}{\gamma
^{2}}-\frac{\beta _{1}^{2}}{\gamma ^{2}}\right) n_{2},  \eqnum{56}
\label{55}
\end{equation}
The last term of Eq. (56) should be negligible compared with the signal,
which requires 
\begin{equation}
\left| \beta _{2}^{2}-\beta _{1}^{2}\right| <\frac{\gamma ^{2}}{\left\langle
n_{2}\right\rangle }.  \eqnum{57}  \label{56}
\end{equation}
To consider the decoherence effect of the light leakage, we define a similar
measuring operator $X_{T}^{\left( \alpha \right) }$ for the leaked light
(55) 
\begin{eqnarray}
X_{T}^{\left( \alpha \right) } &=&\frac{1}{T}\int_{0}^{T}\frac{1}{\sqrt{2}}%
\left[ c_{o\alpha }\left( t\right) +c_{o\alpha }^{\dagger }\left( t\right) %
\right] dt  \nonumber \\
&\approx &\frac{8\sqrt{2}\left| g\right| \chi \sqrt{\beta _{\alpha }}\left(
\alpha -1\right) }{\gamma }\left( n_{1}+n_{2}\right) +\frac{1}{\sqrt{T}}%
X_{T}^{\left( c_{\alpha }\right) }  \eqnum{58}  \label{57} \\
&&-\frac{4\sqrt{2}\left| g\right| \chi \sqrt{\beta _{\alpha }}}{\gamma }%
n_{\alpha },\text{ }\left( \alpha =1,2\right) ,  \nonumber
\end{eqnarray}
where $X_{T}^{\left( c_{\alpha }\right) }$, similar to $X_{T}^{\left(
b\right) }$ defined below Eq. (42), are standard vacuum noise terms. The
last term of Eq. (58) bears some information about $n_{\alpha }$, which
should be completely masked by the vacuum noise term to make the decoherence
effect negligible. This condition requires 
\begin{equation}
\frac{4\sqrt{2}\left| g\right| \chi \left\langle n_{\alpha }\right\rangle }{%
\gamma }\sqrt{\beta _{\alpha }}<\frac{1}{\sqrt{2T}}.  \eqnum{59}  \label{58}
\end{equation}
On the other hand, the measuring time $T$ is bounded from below by Eq. (45),
which, combined with Eq. (59), yields the following requirement for the
leakage rates 
\begin{equation}
\beta _{\alpha }<\frac{\gamma }{\left\langle n_{\alpha }\right\rangle ^{2}},%
\text{ }\left( \alpha =1,2\right) .  \eqnum{60}  \label{59}
\end{equation}
Obviously, this is a much stronger requirement than that given by Eq. (57).

We should mention that there is another kind of absorption by the Kerr
medium, the absorption rate of which is proportional to the cavity photon
number $n_{\alpha }$. This kind of absorption, usually termed two-photon
absorption, cannot be described by Eq. (54). To incorporate the two-photon
absorption, we add an imaginary part to the cross phase modulation
coefficient $\chi $, i.e., $\chi $ is replaced by $\chi +i\chi _{i}$, where $%
\chi _{i}$ describes the two-photon absorption rate. The two-photon
absorption should be negligible compared with the cross Kerr interaction,
which requires $\chi _{i}<\frac{\chi }{\left\langle n_{\alpha }\right\rangle 
},$ $\left( \alpha =1,2\right) .$

\subsection{Self-phase modulation effects}

Normally, a Kerr medium also induces self-phase modulation effects. However,
by a suitable choice of the resonance condition for the Kerr medium, the
self-phase modulation effects can be made much smaller than the cross-phase
modulation \cite{24}, then the self phase modulation interaction is
basically negligible. Here, for completeness, we still calculate the
influence of self-phase modulations. In fact, self-phase modulation of the
ring cavity modes have no influence on the QND measurement. This modulation
adds a term like $-i\chi _{s}b_{i}^{\dagger }b_{i}b_{i}$ $\left(
i=1,2\right) $ in the Langevin equation (38), where $\chi _{s}$ denotes the
self-phase modulation coefficient for the ring cavity modes. We know that
the ring cavity modes $b_{1}$ and $b_{2}$ are in steady states under
adiabatic elimination, and to a good approximation $b_{i}^{\dagger }b_{i}$
can be replaced by $\left\langle b_{i}^{\dagger }b_{i}\right\rangle =4\left|
g\right| ^{2}$. So the term $-i\chi _{s}b_{i}^{\dagger }b_{i}b_{i}$ simply
induces a constant phase shift for the output field $b_{o2}$, and it can be
easily compensated by choosing the initial phase of the driving field $g$.

Self phase modulation of the cavity modes $a_{1}$ and $a_{2}$ plays a more
subtle role. First, it obviously has no influence on the QND\ measurement of 
$n_{1}+n_{2}$, but it influences the resulting state after the QND
measurement. In the purification scheme for two entangled pairs (described
in section III.A), if there is no self-phase modulation, the state after the
QND measurement is given by Eq. (13); and if the self-phase modulation of
the modes $a_{1}$ and $a_{2}$ is considered, the modulation Hamiltonian $%
\hbar \chi _{s}^{\prime }n_{i}^{2}$ $\left( i=1,2\right) $, in which $\chi
_{s}^{\prime }$ is the corresponding self phase modulation coefficient, will
bring the resulting state into

\begin{equation}
\left| j\right\rangle _{A_{1}A_{2}B_{1}B_{2}}^{\prime }=\frac{1}{\sqrt{1+j}}%
\stackrel{j}{\mathrel{\mathop{\sum }\limits_{n=0}}}e^{i\chi _{s}^{\prime }t%
\left[ n^{2}+\left( j-n\right) ^{2}\right] }\left| n,j-n\right\rangle
_{A_{1}A_{2}}\left| n,j-n\right\rangle _{B_{1}B_{2}},  \eqnum{61}  \label{60}
\end{equation}
where $t$ is the interaction time for the self phase modulation. It is
important to note that the state (61) is still a maximally entangled state
with entanglement $\log \left( j+1\right) $. In this sense, self-phase
modulation effects have no influence on the entanglement purification,
though the resulting state is changed.

\subsection{Imperfect coupling from the first ring cavity to the second ring
cavity}

If the coupling between the two ring cavities is not perfect, the relation $%
b_{i2}=b_{o1}$ is not valid any more, and should be replaced by 
\begin{eqnarray}
b_{i2} &=&\sqrt{\mu }b_{o1}+\sqrt{1-\mu }d_{i},  \nonumber \\
d_{o} &=&\sqrt{\mu }d_{i}+\sqrt{1-\mu }b_{o1},  \eqnum{62}
\end{eqnarray}
where $d_{i}$ is the standard vacuum white noise, and $d_{o}$ represents the
leaked light in the imperfect coupling. The quantity $\mu $ describes the
coupling efficiency. This kind of imperfection is very similar to the light
leakage (or absorption) described in subsection VI.C. The difference is that
the imperfect coupling (62) does not cause any unbalance between the two
ring cavities. The only restriction is that the decoherence effect induced
by it should be negligible, which requires

\begin{equation}
\mu >1-\frac{1}{\left\langle n_{1}\right\rangle ^{2}}.  \eqnum{63}
\label{62}
\end{equation}
Eq. (63) suggests that loss of light from the first to the second ring
cavity should be very small.

\subsection{Detector inefficiency}

The detector efficiency of course cannot attain $1$. For a detector with
efficiency $\nu $, the real measured field $b_{o2}^{\prime }$ has the
following relation with the output of the second ring cavity 
\begin{equation}
b_{o2}^{\prime }=\sqrt{\nu }b_{o2}+\sqrt{1-\nu }e_{i},  \eqnum{64}
\label{63}
\end{equation}
where $e_{i}$ is the standard vacuum white noise. This imperfection is
similar to the imperfect coupling considered in the previous subsection. But
now the leaked light depends only on the operator sum $n_{1}+n_{2}$, and
carries no information about the single cavity photon number $n_{1}$, so it
does not induce any decoherence. The only role played by the detector
inefficiency is that it decreases the signal by a factor $\sqrt{\nu }$, so
Eq. (45) on the restriction of the measuring time is now replaced by 
\begin{equation}
T>\frac{\gamma }{64\nu \left| g\right| ^{2}\chi ^{2}},  \eqnum{65}
\label{65}
\end{equation}
Obviously, the detector inefficiency has no important influence on this QND
measurement scheme.

\section{Summary and discussion}

In summary, we have given a detailed description of the purification
protocol which generates maximally entangled states in a finite dimensional
Hilbert space from two-mode squeezed states or from realistic Gaussian
continuous entangled states. The nonlocal Gaussian continuous entangled
states are generated by feeding two distant cavities with the outputs of the
NOPA. The purification operation is based on a local QND measurement of the
total photon number contained in several cavities. We have extensively
analyzed a cavity scheme to do this QND measurement, and have deduced its
working condition. Furthermore, we have discussed many imperfections
existing in a real experiment, and deduced quantitative requirements for 
the relevant experimental parameters. In Tab. 1, we summarize the working
conditions for the collective QND measurement, 
including the requirements for many types of imperfections.
\newpage 
\begin{table}[h]
\begin{tabular}{||c|cc||}
Measuring time & $\frac{\gamma }{64\left| g\right| ^{2}\chi ^{2}}<T<\frac{1}{%
\kappa \left\langle n_{i}\right\rangle }$ &  \\ \hline
Phase instability & $\delta <\frac{4\chi }{\gamma }$ or $\delta _{t}<\frac{%
1024\left| g\right| ^{2}\chi ^{4}}{\gamma ^{3}}$ &  \\ \hline
Cavity imbalance & $\left| \frac{\chi _{2}\gamma _{1}}{\chi _{1}\gamma _{2}}%
-1\right| <\frac{1}{\left\langle n_{2}\right\rangle }$ &  \\ \hline
Absorption (leakage) rate & $\beta _{\alpha }<\frac{\gamma }{\left\langle
n_{\alpha }\right\rangle ^{2}},$ $\left( \alpha =1,2\right) $ &  \\ \hline
Coupling efficiency & $\mu >1-\frac{1}{\left\langle n_{1}\right\rangle ^{2}}$
&  \\ \hline
Detector efficiency & $\nu >\frac{\gamma }{64\left| g\right| ^{2}\chi ^{2}T}$
& 
\end{tabular}
\caption{List of requirements for the QND measurement}
\end{table}
To realize the QND measurement, basically we need high finesse optical
cavities and strong cross Kerr interaction media. A good example for the
strong cross Kerr interaction is provided by the resonantly enhanced Kerr
nonlinearity, which has been predicted theoretically \cite{24,25} and
demonstrated in recent experiments \cite{26}. In those works, the Kerr
medium is a low density cold trapped atomic gas, whose relevant energy level
structure is represented by the four-state diagram shown in Fig. 6 with $%
\left| 1\right\rangle $ being the ground state. The ring cavity mode $b_{i}$
with frequency $\omega _{b}$ is assumed to be resonant with the $\left|
1\right\rangle \rightarrow \left| 3\right\rangle $ transition, and the
cavity mode $a_{i}$ with frequency $\omega _{a}$ ($\omega _{a}$ is quite
different from $\omega _{b}$) is coupled to the $\left| 2\right\rangle
\rightarrow \left| 4\right\rangle $ transition, but with a large detuning $%
\Delta _{42}$. A nonperturbative classical coupling field with frequency $%
\omega _{c}$ resonant with the $\left| 2\right\rangle \rightarrow \left|
3\right\rangle $ transition creates an electromagnetically induced
transparency (EIT) for the cavity fields $a_{i}$ and $b_{i}$. 
\begin{figure}[tbp]
\epsfig{file=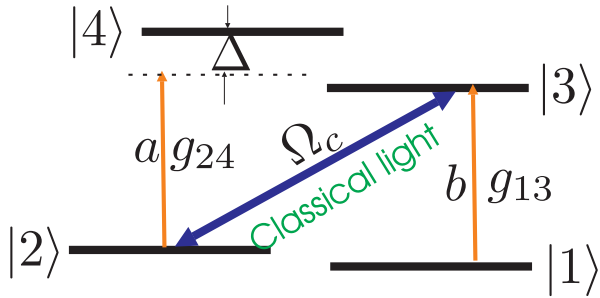,width=8cm}
\caption{Level structure of the atoms. }
\end{figure}
In this configuration, the one-photon absorption of the medium is eliminated
due to quantum interference, and the cross Kerr nonlinearity is only limited
by the two-photon absorption (the self Kerr nonlinearity is negligible
provided that $\left| \omega _{a}-\omega _{b}\right| \gg \Delta _{42}$).
After adiabatically eliminating all the atomic levels, the cross phase
modulation coefficient is given by \cite{24} 
\begin{equation}
\chi \sim \frac{3\left| g_{13}\right| ^{2}\left| g_{24}\right| ^{2}}{\Omega
_{c}^{2}\Delta _{42}}n_{\text{atom}},  \eqnum{66}
\end{equation}
where $g_{24}$ and $g_{13}$ are the coupling coefficients between the atoms
and the cavity modes $a_{i}$ and $b_{i}$, respectively, $\Omega _{c}$
denotes the Rabi frequency of the coupling field, and $n_{\text{atom}}$ is
the total atom number contained in the cavity. The two-photon absorption
rate $\chi _{i}$ is connected with $\chi $ by the relation $\chi _{i}/\chi
=\gamma _{42}/\Delta _{42}$, where $2\gamma _{42}$ is the spontaneous
emission rate from level $\left| 4\right\rangle $ to level $\left|
2\right\rangle $. To justify the adiabatic elimination, one requires that $%
\frac{\left| g_{13}\right| ^{2}n_{\text{atom}}}{\Omega _{c}^{2}}<1$ \cite
{27,28}. As an estimation, if one takes $\frac{\left| g_{13}\right| ^{2}n_{%
\text{atom}}}{\Omega _{c}^{2}}\sim 0.2,$ $g_{24}/2\pi \sim 10$MHz, $\gamma
_{42}/2\pi \sim 30$MHz, and $\Delta _{42}\sim 10\gamma _{42}$, the
coefficient $\chi $ is about $\chi /2\pi \sim 0.2$MHz, and the two-photon
absorption rate $\chi _{i}\sim 0.1\chi $. This value of the cross phase
modulation coefficient $\chi $ is not large enough to realize a
single-photon turnstile device \cite{24}, but it is enough for performing
QND measurements of the photon number. For example, if the mean photon
number $\left\langle n_{1}\right\rangle =\left\langle n_{2}\right\rangle
=\sinh ^{2}\left( r\right) \sim 1.4$ with the squeezing parameter $r\sim 1.0$%
, we choose the decay rates $\kappa /2\pi \sim 4$MHz and $\gamma /2\pi \sim
100$MHz (these values for decay rates are obtainable in current experiments
), and let $g\sim 50$ (for a cavity with cross area $S\sim 0.5\times 10^{-4}$%
cm$^{2}$, $g\sim 50$ corresponds to a coherent driving light with intensity
about $10$mWcm$^{-2}$). With the above parameters, all the requirements
listed in Tab. 1 can be satisfied if we choose the measuring time $T\sim 8$%
ns. Note that the light speed can be much reduced in the EIT medium \cite{26}%
, so it is possible to get a reduced cavity decay rate $\kappa $ with the
same finesse mirrors, and then more favorable parameters can be given for
the QND measurement. Note also that a large Kerr nonlinearity based on EIT
can also be obtained in other systems, such as trapping a single atom in a
high finesse cavity \cite{PW}. So the example discussed here is not the
unique choice.

We thank P. Grangier and S. Parkins for discussions. This work was supported
by the Austrian Science Foundation, by the European TMR network Quantum
Information, and by the Institute for Quantum Information. GG acknowledges
support by the Friedrich-Naumann-Stiftung.

\end{document}